\documentclass[prb,aps,twocolumn,showpacs,floats]{revtex4}
\usepackage{psfig}

\begin{document}


\title{Charge-ordered ferromagnetic phase in manganites}

\author{Tran Minh-Tien}

\affiliation{Institute of Physics, NCST,  
P.O. Box 429, Boho, 10000 Hanoi, Vietnam, \\
Electrophysics Department, National Chiao Tung University, Hsinchu 300, Taiwan.
}



\begin{abstract}
A mechanism for
charge-ordered ferromagnetic phase in manganites is proposed.
The mechanism  is
based on the double exchange in the presence of diagonal disorder.
It is modeled by a
combination of the Ising double-exchange  and the 
Falicov-Kimball model. 
Within the dynamical
mean-field theory the charge and spin correlation function are
explicitely calculated. It is shown that 
the system exhibits two successive phase transitions. The first one is
the ferromagnetic phase transition, and the second one is
a charge ordering. As a result a charge-ordered ferromagnetic phase
is stabilized at low temperature.

\end{abstract} 



\maketitle

There has been much recent interest in the properties of
doped manganese oxides R$_{1-x}$A$_x$MnO$_3$ 
(R=rare earth,  A=Ca,Sr).
\cite{salamon,dagotto}
These materials present a very rich phase diagram  involving phases with spin, 
charge and orbital order. 
The physically relevant electrons in manganites are those from
the Mn $3d$ levels, which are split by the cubic crystal field into
triply degenerate $t_{2g}$ levels and higher-energy doubly degenerate
$e_g$ levels. Electrons from the $e_g$ levels are able to hop between
Mn sites  and form  a conduction band. Electrons from the $t_{2g}$ levels
are localized. The itinerant electrons and local spins are correlated
by the double-exchange (DE) mechanism. {\cite{zener,anderson}}
The main feature of the DE is a cooperative effect where
the motion of an itinerant electron favors the ferromagnetic (FM)
ordering of local spins and, vice versa, the presence of the FM
order facilitates the motion of the itinerant electron. The
DE model qualitatively describes some of the magnetic properties
of manganites, \cite{dagotto,furukawa}  and provides
a well-established starting 
point toward comprehensive
understanding of the phase diagram of  manganites.

Recently experiments have shown that beside the FM order 
a charge order can exist in the manganites. \cite{loudon,uehara} The 
charge order exists in regions with no net magnetization and, surprisingly, can also
occur in FM regions. \cite{loudon} 
Doping of A$^{2+}$ ions creates Mn$^{4+}$ holes in a Mn$^{3+}$ background.
The presence of two valence states Mn$^{3+}$ and Mn$^{4+}$ may lead the 
compounds to a charge-ordered (CO) state for appropriate doping.
However,  the DE model alone cannot explain the CO state which coexists in the
FM phase. 
In principle, the nearest-neighbor Coulomb repulsion may stabilize a CO state.
However, a large nearest-neighbor repulsion likely destabilizes the 
homogeneous FM 
state and may produce a checkerboard charge order in three directions. \cite{dagotto}
Another possible mechanism for the CO phase stabilization  is
the coupling of itinerant electrons to the Jahn-Teller distortions. 
However, the electron Jahn-Teller phonon coupling
can only stabilize a CO-FM state where the CO phase transition occurs 
before the FM transition. \cite{yunoki} 
At half filling experiments have only observed a charge order below the 
FM transition temperature. \cite{loudon,dagotto} Therefore the Jahn-Teller coupling
is unlikely responsible for the appearance of the CO-FM state at least at half filling.
In this paper we present a possible alternative explanation  for the CO-FM state in
the manganites. The key idea is an interplay of the DE and randomness of the 
A-site substitution.
The randomness  is inevitably introduced
by experiments.  The
importance of the randomness has  been discussed both 
experimentally and theoretically. \cite{salamon,dagotto}
The randomness can substantially decrease the critical temperature of 
the FM
transition. {\cite{zhong,letfulov,kogan}} 
Here we will incorporate the randomness of A-site substitution into the DE
model.
For simplicity, we adopt the randomness by  A-site substitution
as a random local potential of the itinerant electrons, although the
randomness may cause other effects, for instance, randomness of the hoping
or exchange integral. \cite{dogoto1}
It is
well known that the diagonal disorder with binary distribution can be
modeled by the Falicov-Kimball (FK) model. {\cite{falicov}} Although the FK
model is simple, it contains a rich variety of phases.
In particular, it illustrates the disorder-order phase transition 
driven by electron interaction. \cite{kennedy,brandt1}
Incorporating the diagonal
disorder of the FK type into the DE model,
one may expect
that a disorder-order phase transition could present. 
When the phase transition occurs, a CO-FM phase may be stabilized at 
low temperature.  
  In order to
detect the phase transition we study the charge and spin response
of system by using the dynamical mean-field theory (DMFT). {\cite{GKKR}}
The DMFT has extensively been used  
for investigating strongly
correlated electron systems. {\cite{GKKR}}  
Within the DMFT we explicitely calculate
the charge and spin correlation function. 
We find that the system stabilizes a CO-FM
state at low temperature. 

The system which we study is described by the following
Hamiltonian
\begin{eqnarray}
H &=& - \frac{t}{\sqrt{d}} 
\sum_{<ij>,\sigma} c^{\dagger}_{i\sigma} c^{\null}_{j\sigma}
- \mu \sum_{i\sigma} n_{i\sigma} - 2 J_{H} \sum_{i} S^{z}_{i} s^{z}_{i} 
+ 
\nonumber \\
&&
E_{w} \sum_{i} w_{i} + 
U \sum_{i\sigma} n_{i\sigma} w_{i} ,
\label{hamil}
\end{eqnarray}
where $c^{\dagger}_{i\sigma}$($c^{\null}_{i\sigma}$) is the creation
(annihilation) operator of an itinerant electron with spin $\sigma$
at lattice site $i$; $t/\sqrt{d}$ is the hoping parameter of the itinerant
electrons. Here we have rescaled the hoping parameter with the dimension
$d$ of the system. 
$S^{z}_{i}$ is the $z$ component of local
spin at lattice site $i$, and for simplicity, it takes two values $\pm 1$. 
$s^{z}_{i}=(n_{i\uparrow}-n_{i\downarrow})/2$,
$n_{i\sigma}=c^{\dagger}_{i\sigma} c^{\null}_{i\sigma}$, $w_{i}$ is a
classical variable that assumes the value $1$($0$) if site $i$ is occupied
(not occupied) by A ion. $U$ is the disorder strength and is mapped
onto the difference in the local potential which splits 
energetically favor of Mn$^{3+}$ and Mn$^{4+}$ ions. 
The expectation value $x=\sum_{i} \langle w_i \rangle / N$, ($N$ is the number of lattice sites), corresponds to
the concentration of unfavorable Mn$^{4+}$ sites. 
The chemical potential $\mu$ controls
the carrier doping, while $E_{w}$ controls the fraction of the sites
having the additional local potential. We shall use the condition
$n + x =1 $, where $n=\sum_{i\sigma} \langle n_{i\sigma} 
\rangle /N$ is the electron doping. 
This condition determines $E_w$ for
each doping $n$. The third term of Hamiltonian (\ref{hamil}) is the Hund coupling
of itinerant and local electrons. 
For simplicity we only take into account
the Ising part of the Hund coupling. This simplification does not allow 
any spin-flip processes,
which can be important at low temperature where spin-wave excitations
may govern the thermodynamics of the system.
 However,  in the DE processes the spin of  itinerant
electron ferromagnetically aligns with the local spin, hence, the Ising part
of the Hund interaction plays a dominant role.
The DMFT calculations for the DE model with classical local spins show 
that the simplification of the Hund coupling does not change the self energy
of the single-particle Green function. {\cite{furukawa}}
Moreover, within  the DMFT the numerical results for
quantum local spins do not show a significant difference from the ones for
classical local spins. {\cite{nagai}}
Thus, one expects that within the DMFT the simplification of the
Hund interaction does not result in a serious backwardness.
$J_H$ is the strength of the Hund
coupling, and in the following we will take the limit $J_H \rightarrow \infty$.
The first three terms of Hamiltonian (\ref{hamil}) constitute a 
simplified DE model. This simplified model captures the most essential
ingredient of the DE processes.
The last two terms of Hamiltonian (\ref{hamil}) 
describe a binary randomness of the A-site substitution. They together
with the hoping term form the FK model.
{\cite{falicov}}
It is well known that within the FK model the $U$ term induces
a disorder-order phase transition. \cite{kennedy,brandt1} 
At low temperature a checkerboard
ordering phase is stabilized. Hence, the model (\ref{hamil})
may display an interplay of the FM and CO phase.

We solve model (\ref{hamil})
by the DMFT.
The DMFT is based on the infinite-dimension limit.
In the infinite-dimension limit  the self energy is pure local and has 
no momentum
dependence. The Green function of itinerant electrons satisfies
the Dyson equation
\begin{equation}
G_{\sigma}({\bf k}, i \omega_n) = 
\frac{1}{i \omega_n - \varepsilon({\bf k}) + \mu - 
\Sigma_{\sigma}(i \omega_n)} ,
\label{greenlattice}
\end{equation}
where $\omega_n=\pi T (2 n + 1)$, 
$\varepsilon({\bf k})= - 2 t\sum_{j=1}^{d} \cos(k_j)$, 
and $\Sigma_\sigma(i \omega_n)$ is the self energy. 
In the infinite-dimension limit the
bare density of states of itinerant electrons becomes
$
\rho(\varepsilon)= 
\exp(-\varepsilon^2/{t}^2) / \sqrt{\pi} t
$
and we take $t$ as the unit of energy ($t=1$).
The self energy is determined by solving an effective
single-site problem. The effective action of this
problem is
\begin{eqnarray}
S_{\text{eff}}  =  
\sum_\sigma \int  d\tau d\tau' c_\sigma^\dagger(\tau) 
\big( \frac{\partial}{\partial \tau} \delta(\tau-\tau')
+ \Lambda_\sigma(\tau-\tau') \big) c_\sigma(\tau') 
 \nonumber \\
 + \sum_\sigma \int d\tau  
c_\sigma^\dagger(\tau) \big(
- \mu + U w - J_{H} \sigma S^z \big) c_\sigma(\tau)
+ E_w w \mbox{\hspace{0cm}} , 
\nonumber 
\label{action}
\end{eqnarray}
where $\Lambda_\sigma(\tau)$ describes the effective medium.
This effective single-site problem  can exactly be  solved.
Indeed,
the dynamics of the
localized spin $S^z$ and impurity $w$ involved in the effective action
are independent, hence, we could independently take the trace over $S^z$ and
$w$  in calculating the partition function. This is
similar to the DMFT solving of the FK model. {\cite{brandt}}
We obtain
 the local Green function in the limit $J_H \rightarrow \infty$
\begin{equation}
G_\sigma(i\omega_n) = 
\frac{W_{0\sigma}}{Z_\sigma(i\omega_n) } +
\frac{W_{1\sigma}}{Z_\sigma(i\omega_n) - U} ,
\label{greenlocal1}
\end{equation}
where $Z_\sigma(i\omega_n) =  i\omega_n + \mu - \Lambda_\sigma(i\omega_n)$, and
\begin{eqnarray*}
W_{\alpha\sigma} = \bigg\{
\sum_{\alpha'=0,1} \sum_{\sigma'}  \exp \Big[ -\beta E_w (\alpha'-\alpha) +
\nonumber \\ 
 \sum_n 
\ln \Big( \frac{Z_{\sigma'}(i\omega_n)-\alpha' U}{Z_{\sigma}(i\omega_n)-\alpha U} 
\Big) \Big]  \bigg\}^{-1} 
\end{eqnarray*}
with $\alpha=0,1$.
In taking the limit $J_H \rightarrow \infty$
in  deriving Eq. (\ref{greenlocal1}) we  must first renormalize the chemical
potential  $\mu \rightarrow \mu + J_{\text H}$.
The self energy is 
determined by the Dyson equation for the effective single-site problem
\begin{equation}
\Sigma_\sigma(i\omega_n) = Z_\sigma(i\omega_n)
- G_\sigma^{-1}(i\omega_n) .
\label{selfenergy}
\end{equation}
Within the DMFT, the local Green function must coincide with the 
single-site Green
function of the original lattice, i.e.,
\begin{equation}
G_\sigma(i\omega_n) = \frac{1}{N} \sum_{\bf k} G_\sigma({\bf k},i\omega_n) .
\label{greenlocal} 
\end{equation}  
Eqs. (\ref{greenlattice})-(\ref{greenlocal})
form the complete set of equations, which self consistently determine
the self energy and Green function.

We are interested in calculating  the charge (c) and 
spin (s) correlation
function
\begin{equation}
\chi^{\text{c(s)}}(i,j) = \big\langle
( \delta n_{i\uparrow} \pm \delta n_{i\downarrow} )
( \delta n_{j\uparrow} \pm \delta n_{j\downarrow} ) \big\rangle 
\end{equation}
in the homogeneous paramagnetic (PM) phase 
($\delta n_{i\sigma} = n_{i\sigma} - \langle n_{i\sigma} \rangle$).
In order to calculate the charge and spin response of system
one has to introduce an external field into the Hamiltonian. 
The charge and spin correlation function can   be  obtained by
differentiating the Green function respected to the external field,
and then taking the zero limit of the field. {\cite{brandt}}
Following the standard techniques, {\cite{brandt}} one can express
the correlation functions in the terms of charge (c) and spin (s) susceptibility 
$\chi^{\text{c(s)}}({\bf q},i\omega_n)$
in momentum space
\begin{equation}
\chi^{\text{c(s)}}({\bf q}) = - T^2 \sum_n 
\chi^{\text{c(s)}}({\bf q}, i\omega_n) .
\label{corr}
\end{equation}
The charge and spin susceptibility 
 can be obtained by differentiation of the
Green function. \cite{brandt} We obtain 
\begin{equation}
\chi^{\text{c(s)}}({\bf q},i\omega_n) = 
 \frac{ 2 + \sum\limits_{\alpha=0,1} 
\big(  \frac{\partial \Sigma(i\omega_n)}{\partial 
W_{\alpha}}
\big)_{G,W_{1-\alpha}}
\Gamma^{\text{c(s)}}_{\alpha}({\bf q})}{
[\chi_{0}({\bf q},i\omega_n)]^{-1} -  
\big(  \frac{\partial \Sigma(i\omega_n)}{\partial 
G(i\omega_n)}
\big)_{W} } ,
\label{chi1}
\end{equation}
where 
$\chi_{0}({\bf q},i\omega_n)=\sum_{\bf k} G({\bf k}+{\bf q},i\omega_n)
G({\bf k},i\omega_n)$. 
\begin{figure}[b]
\vspace{-0.5cm}
\centerline{
\psfig{file=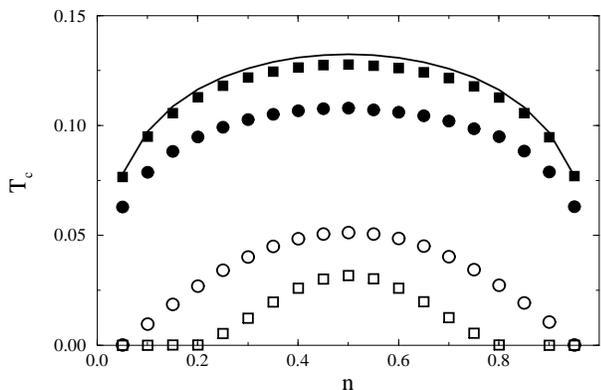,width=0.5\textwidth}}
\vspace{-0.5cm}
\caption{The critical temperature $T_c$ as a function of doping $n$
for $U=0.5$ (squares), $U=1$ (circles). The filled (open) symbols are $T_c$
of the FM (CO) phase transition.
The solid line is $T_c$ of the FM transition without disorder ($U=0$). 
}
\label{fig1}
\end{figure}
The matrix $\widehat{\Gamma}^{\text{c(s)}}({\bf q})$  
satisfies the following equation
\begin{equation}
\widehat{B}^{\text{c(s)}}({\bf q}) \widehat{\Gamma}^{\text{c(s)}}({\bf q})
= \widehat{Q}^{\text{c(s)}}({\bf q}) ,
\label{b}
\end{equation}
where $ \widehat{B}^{\text{c(s)}}({\bf q}) $ and
$\widehat{Q}^{\text{c(s)}}({\bf q}) $ have the following elements
\begin{eqnarray}
B_{\alpha \beta}^{\text{c(s)}}({\bf q}) = \delta_{\alpha \beta} + 
\mbox{\hspace{5cm}}
\nonumber \\
 \sum_{n} 
\frac{ A^{\text{c(s)}}_\alpha(i\omega_n) \eta({\bf q},i\omega_n)
G(i\omega_n) 
\big( \frac{\partial \Sigma(i\omega_n)}{\partial W_{\beta}} 
\big)_{G,W_{1-\beta}}
}{1 - G^{2}(i\omega_n) 
\big(
 \frac{\partial \Sigma(i\omega_n)}{\partial G(i\omega_n) } 
\big)_{W}
+  \eta({\bf q},i\omega_n) G(i\omega_n)
} , \\
Q_{\alpha}^{\text{c(s)}}({\bf q}) = 
\mbox{\hspace{6cm}}
\nonumber \\
\sum_n \frac{ A^{\text{c(s)}}_\alpha(i\omega_n) \Big(
G^{2}(i\omega_n) \big(
 \frac{\partial \Sigma(i\omega_n)}{\partial G(i\omega_n) } 
\big)_{W} - 1 \Big) }{
1 - G^{2}(i\omega_n) 
\big(
 \frac{\partial \Sigma(i\omega_n)}{\partial G(i\omega_n) } 
\big)_{W}
+  \eta({\bf q},i\omega_n) G(i\omega_n)
} 
\label{q}
\end{eqnarray}
with $\alpha,\beta=0,1$.
In deriving Eqs.~(\ref{b})-(\ref{q}) we have used the standard conversion \cite{brandt} 
$
[ \chi_{0}({\bf q}, i\omega_n) ]^{-1} =
[ G_\sigma(i\omega_n) ]^{-2} + \eta({\bf q},i\omega_n)
[G_\sigma(i\omega_n) ]^{-1}, 
$ 
and introduced quantity
$
A^{\text{c(s)}}_{\alpha}(i\omega_n) =
\delta W_{\alpha \uparrow}/ \delta Z_{\uparrow}(i\omega_n) \pm
\delta W_{\alpha \uparrow}/ \delta Z_{\downarrow}(i\omega_n) .
$
In the infinite dimension limit all of the wave vector dependence
of $\chi_{0}({\bf q},i\omega_n)$ and $\eta({\bf q},i\omega_n)$ 
include in the term $X({\bf q})=\sum_{j=1}^{d} \cos q_j /d$.
Hence, the spin and charge correlation function only depend on momentum
 via $X({\bf q})$.
Each of the derivatives appearing in Eqs.~(\ref{chi1})-(\ref{q}) can directly be
 calculated from the DMFT solution of Eqs.~(\ref{greenlattice})-(\ref{greenlocal}).
In such the way  $\widehat{B}^{\text{c(s)}}({\bf q})$ and 
$\widehat{Q}^{\text{c(s)}}({\bf q})$ are calculable once the
self-consistent equations of the DMFT are solved. 
Equation~(\ref{b})
reveals that $\widehat{\Gamma}^{\text{c(s)}}({\bf q})$ will diverge
at a temperature where the determinant of 
$\widehat{B}^{\text{c(s)}}({\bf q})$ vanishes, while 
$\widehat{Q}^{\text{c(s)}}({\bf q})$ remains finite. This results
in an unphysical change of the sign of corresponding correlation
function $\chi^{\text{c(s)}}({\bf q})$ so that the assumption of
the homogeneous PM phase fails for lower temperature.
By a similar way one could also calculate the spin correlation
function $\chi_{S}({\bf q})$ of  local spins. After some calculations we obtain
\begin{eqnarray}
\chi_{S}({\bf q}) = \gamma_{0}({\bf q}) + \gamma_1({\bf q}) , 
\label{chis}\\
\sum_{\beta=0,1} B^{\text{s}}_{\alpha \beta}({\bf q})  \gamma_\beta({\bf q}) =
\frac{2 W_{\alpha}}{T} . 
\label{gs}
\end{eqnarray}
From Eqs.~(\ref{b}), (\ref{chis})-(\ref{gs}) one can see that the spin correlation
function of itinerant electrons and local spins will
diverge at the same temperature where the 
determinant of 
$\widehat{B}^{\text{s}}({\bf q})$ vanishes. 
This means that the spin of itinerant electrons  parallel aligns  with local spin, and thus
is an important feature of the DE.
\begin{figure}[t]
\centerline{
\psfig{file=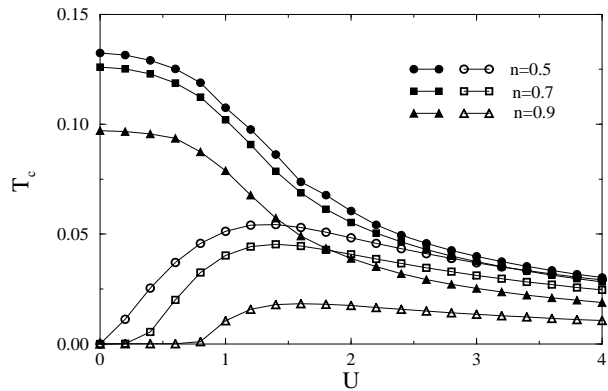,width=0.5\textwidth} }
\vspace{-0.5cm}
\caption{
The critical temperature $T_c$ as a function of  $U$ for various doping $n$.
The filled (open) symbols are $T_c$
of the FM (CO) phase
transition.}
\label{fig2}
\end{figure}
We calculate the charge and spin correlation function 
(\ref{corr}) by solving the DMFT set of self-consistent 
Eqs.~(\ref{greenlattice})-(\ref{greenlocal}). 
We are only interested in the FM and checkerboard CO phase stability. 
Hence, we only calculate the spin correlation  function at 
$X_{\bf q}=1$ and the charge correlation function at $X_{\bf q}=-1$. It is found that
the spin correlation function $\chi^{\text{s}}(X_{\bf q}=1)$ always diverges at a 
critical temperature. This is the signal of the FM phase transition. The charge correlation
function $\chi^{\text{c}}(X_{\bf q}=-1)$ only diverges for $U\not=0$. This means that
without the disorder the system  always is homogeneous. In Figs.~\ref{fig1} and 
\ref{fig2}    we present
the critical temperature $T_c$ of the FM and CO phase transition as a function of doping 
and 
disorder strength. At half filling $n=0.5$ both critical temperatures reach their
maximal value. The $T_c$ of the FM transition always decreases with increasing disorder
strength. This means that the disorder substantially decrease $T_c$ of 
the FM transition. \cite{zhong,letfulov,kogan}
At the same time, with increasing $U$, $T_c$ of the CO phase transition first increases,
reaches its maximal value, and then decreases.  The behavior of $T_c$ of the CO
phase transition
is similar to the one in the FK model. \cite{brandt} At very strong disorder
($U \gg 1$) the two critical temperatures approach to a same value .  
One also notices that $T_c$ of the
CO phase transition always is smaller than the FM transition temperature. 
Thus, one may expect that the CO
state is stabilized in the FM phase at low temperature.
 However, this CO phase stability is respected to the homogeneous PM
phase, and for safety we also study  an inhomogeneous phase. We divide the lattice
into two penetrating sublattices $A$ and $B$. This lattice division allows us to
study the checkerboard CO phase. 
\begin{figure}[t]
\centerline{ 
\psfig{file=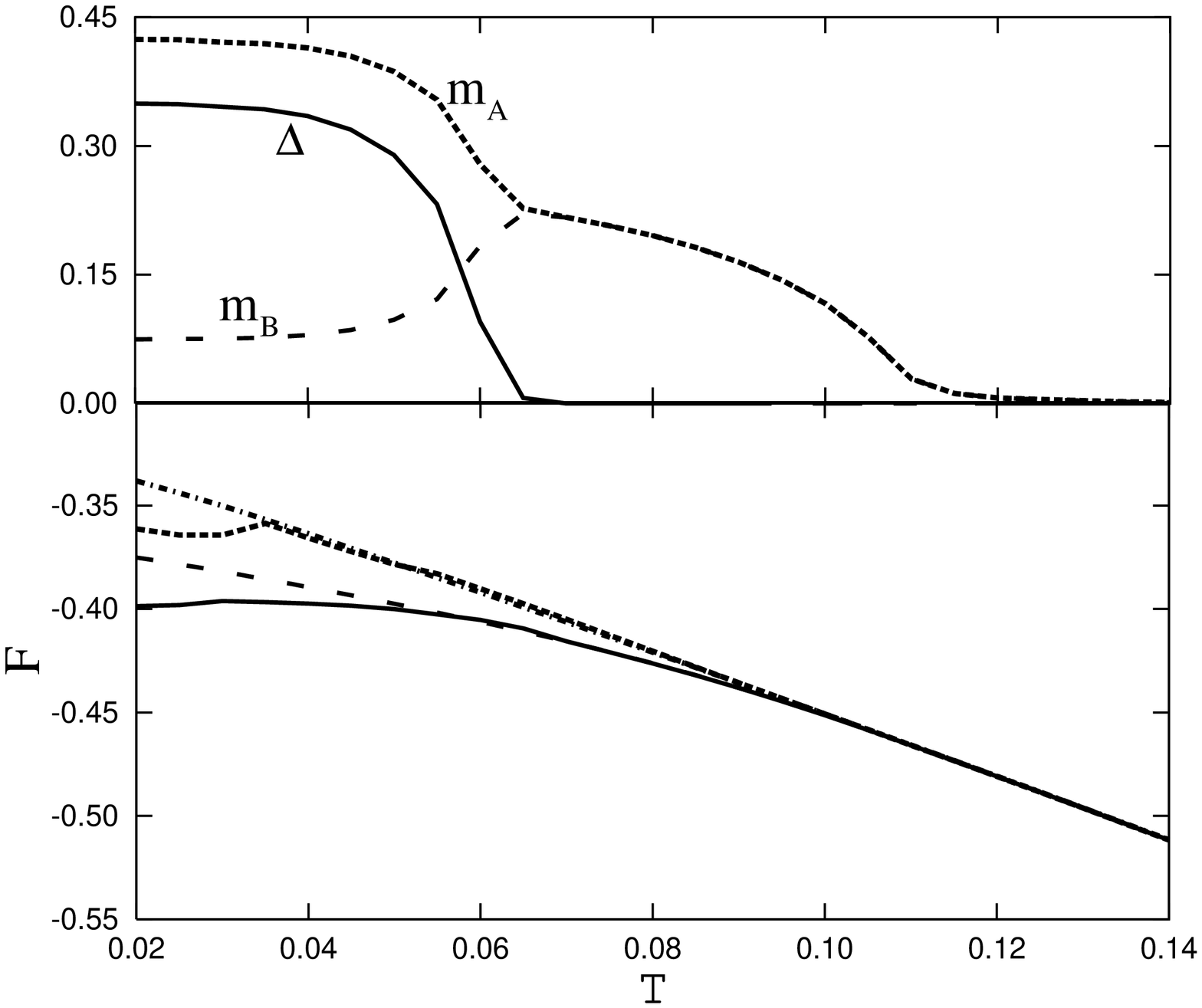,width=0.5\textwidth,angle=-90} }
\caption{
Upper panel: the temperature dependence of the spin magnetization $m_A$ 
($m_B$)  of sublattice A(B)
(the dot and dashed line), and of the charge-order parameter $\Delta$ (the 
solid line). \\
Lower panel: the temperature dependence of the free 
energy F. 
 The solid, long dashed, short dashed, and dot-dashed line are the free energy in the
CO-FM,  homogeneous FM, CO-PM, and
homogeneous PM phase, respectively ($U=1$, $\mu=U/2$).  
}
\label{fig3}
\end{figure}
By using the standard technique \cite{GKKR}
the matrix Green function can be written in the following form
\begin{eqnarray*}
\widehat{G}_{\sigma}^{-1}({\bf k},i\omega_n) =
\left(\begin{array}{cc}
i\omega_n + \mu - \Sigma_{\sigma}^{A}(i\omega_n)  & -\varepsilon({\bf k}) \\
-\varepsilon({\bf k}) & i\omega_n + \mu - \Sigma_{\sigma}^{B}(i\omega_n) 
\end{array} \right) ,
\end{eqnarray*}
where $\Sigma_{\sigma}^{A(B)}(i\omega_n)$ is the self energy of the Green function
of sublattice $A$($B$). The self energies are determined by solving the effective
problem  of single site of the sublattices. \cite{GKKR}
We find that at low temperature a 
checkerboard CO-FM  state is stabilized.  We plot the magnetization 
$m_{A(B)}=2 \sum_{i\in A(B)} \langle s^{z}_i \rangle / N $
of sublattice $A$ ($B$)
as a function of temperature in Fig.~\ref{fig3} (upper panel). In this figure 
we also plot
the temperature dependence of the charge-order parameter 
$\Delta =( \sum_{i\in A,\sigma}   \langle n_{i\sigma} \rangle - 
\sum_{j\in B,\sigma}   
\langle n_{j\sigma} \rangle )/N$.
It shows that
below  a critical temperature the magnetizations of both sublattices exist.
They equal to each other until another critical temperature, where the
charge-order parameter exists.
At low temperature the system is in the checkerboard CO-FM
state.  In this phase the charge order coexists in the FM state, as  
experimentally observed. \cite{loudon} 
In the way the system exhibits two successive
phase transitions. Initially the system goes to the homogeneous FM phase, and after that
to the checkerboard CO phase, that the CO-FM phase is stabilized.
We also calculate the free energy of the system. The free energy
can only be  expressed in terms of  local quantities. \cite{GKKR}
We plot the temperature dependence of the free energy $F$ in Fig~\ref{fig3}
(lower panel). It shows that the CO-FM state has lowest free energy, hence
the state must be stabilized at low temperature. 

In conclusions, we have  proposed a mechanism  for the CO-FM phase which has 
recently been observed. The mechanism  is based on a combination of diagonal disorder
and a simple DE model with local Ising spins. Employing the DMFT we have calculated
the charge and spin correlation function. It is found that the FM and CO state are 
stabilized at low temperature. As a result the checkerboard charge order can occur
in the FM state.   
However, the manganites
are too complicated a system to be  completely described  by this simple model. 
In particular, the phase with inhomogeneous percolation of  FM and CO regions
is beyond the scope of this paper. 

This work was supported by the National Program of Basic Research 
on Natural Science of Vietnam, Project 4.1.2. The writing was completed at the National
Chiao Tung University, and was supported by Project NSC 91-2811-M-009-006
of ROC.

\end{document}